# Critical Nucleus Size, Composition, and Energetics in the Synthesis of Doped ZnO Quantum Dots

*J. Daniel Bryan, Dana A. Schwartz, and Daniel R. Gamelin*

**Department of Chemistry, University of Washington, Seattle, WA 98195-1700**

**Abstract**

The influence of $Co^{2+}$ ions on the homogeneous nucleation of ZnO is examined. Using electronic absorption spectroscopy as a dopant-specific in situ spectroscopic probe, $Co^{2+}$ ions are found to be quantitatively excluded from the ZnO critical nuclei but incorporated nearly statistically in the subsequent growth layers, resulting in crystallites with pure ZnO cores and $Zn_{1-x}Co_xO$ shells. Strong inhibition of ZnO nucleation by $Co^{2+}$ ions is also observed. These results are explained using the classical nucleation model. Statistical analysis of nucleation inhibition data allows estimation of the critical nucleus size as $25 \pm 4$ $Zn^{2+}$ ions. Bulk calorimetric data allow the activation barrier for ZnO nucleation containing a single $Co^{2+}$ impurity to be estimated as 5.75 kcal/mol cluster greater than that of pure ZnO, corresponding to a $1.5 \times 10^4$-fold reduction in the ZnO nucleation rate constant upon introduction of a single $Co^{2+}$ impurity. These data and analysis offer a rare view into the role of composition in homogeneous nucleation processes, and specifically address recent experiments targeting formation of semiconductor quantum dots containing single magnetic impurity ions at their precise centers.

**Introduction**

Doped semiconductor nanocrystals[1] have recently become the subject of an active area of research, motivated by the prospect of combining the attractive properties of quantum dots (e.g., quantum confinement and high luminescence quantum yields) with those of doped semiconductors (e.g., controlled carrier concentrations, sub-band gap optical transitions, and magnetism). Magnetically doped semiconductor nanostructures are beginning to attract particular attention in the area of spin-based electronics, or "spintronics".[2] Bulk semiconductors doped with magnetic impurities, generally referred to as "diluted magnetic semiconductors" (DMSs, or sometimes "semimagnetic semiconductors"), have been studied extensively since recognition that their excitonic Zeeman splittings typically exceed those of the corresponding non-magnetic (undoped) semiconductors by ca. 2 orders of magnitude.[3,4] These so-called "giant Zeeman splittings" make possible new applications of DMSs in magneto-optical devices[5] or as spin polarization components in spin-dependent electroluminescent[2, 6] or tunnel junction[7] devices. Many recently proposed or tested spintronics devices involve nanoscale DMSs as key functional components.[6, 8]

Precise control over the dopant speciation is expected to be critical for the success of nanoscale DMS device applications. Molecular beam epitaxy (MBE) has been demonstrated as an excellent method for growing high-quality doped nanostructures on surfaces or within complex device heterostructures.[2,6,9] In contrast, relatively little is known about doping semiconductor nanocrystals using direct chemical methods. Only recently have chemists begun to develop solution-phase approaches to the preparation of nanoscale DMSs. Direct chemical approaches offer unparalleled processing opportunities. The product DMS nanocrystals can be purified by standard chemical techniques, and dopant speciation can be improved by stripping off surface dopants or by isocrystalline core/shell growth.[10-13] Direct chemical approaches are generally compatible with large-scale production, and the resulting colloidal suspensions are ready for application in self-assembly strategies that have become a mainstay of nanotechnology.[14] The incorporation of dopants into semiconductor nanocrystals by chemical methods remains a practical challenge, however, since impurities generally tend to be excluded during crystal growth. Indeed, recrystallization and zone refinement are both widely used methods of purification. The elevated synthesis temperatures required for many semiconductor nanocrystals may exacerbate this problem by providing additional thermal energy for dopant migration.[10,15]

Dopant location *within* semiconductor nanocrystals may also be an important consideration for some applications because the spatial probability distributions of the band electrons are confined by the nanocrystal boundaries.[16] For example, the wavefunction of the exciton of a spherical quantum dot has its highest probability density precisely at the center of the nanocrystal. Because the giant excitonic Zeeman splitting in a doped quantum dot depends on the spatial overlap of the excitonic and dopant wavefunctions, magnetic dopants will therefore generate the greatest excitonic Zeeman splittings when placed at the precise centers of the quantum dots.[17] As emphasized previously,[17,18] in the limit of a single dopant ion per quantum dot the giant Zeeman splitting will persist even in the absence of an applied magnetic field. For this reason, several recent articles have reported experiments in which the target materials were colloidal semiconductor quantum dots each containing a single magnetic impurity ion, preferably located at the precise center of the nanocrystal.[10,15,17]

　　　Recently, we reported the observation that cobalt ions are selectively excluded from the initial nucleation events during the synthesis of $Co^{2+}$:ZnO DMS quantum dots grown from homogeneous solutions, despite the high solid solubility of $Co^{2+}$ found during the growth of nanocrystalline $Co^{2+}$:ZnO.[13] In this paper, we analyze electronic absorption spectroscopic data collected during the synthesis of $Co^{2+}$:ZnO DMS quantum dots using a classical nucleation model to understand the important influence dopants have on spontaneous crystal nucleation. We discuss the microscopic origins of the observed nucleation inhibition and use a statistical model to deduce the size of the critical nucleus for ZnO under these conditions, i.e. the size of the smallest ZnO crystallite that may survive to grow into an observable nanocrystal. From this analysis, nucleation of ZnO containing a single $Co^{2+}$ impurity is determined to be $10^4$ times slower than nucleation of pure ZnO. These results provide a clear demonstration that nucleation processes may be highly sensitive to even small perturbations, and as such pertain directly to the aforementioned objective of defect incorporation at the precise centers of quantum dots grown from solution.

**Experimental and Results**

　　　The procedure for synthesis of $Co^{2+}$:ZnO has been described in detail previously.[13] Ethanolic solutions of $N(CH_3)_4OH$ were added to solutions of $Zn(OAc)_2 \bullet 2H_2O$ and $Co(OAc)_2 \bullet 4H_2O$ dissolved in DMSO and stirred for several minutes at room temperature until pseudo-equilibrium conditions were reached. Reaction progress was monitored using electronic absorption spectroscopy. High-resolution electron microscopy images were collected at Pacific Northwest National Laboratory using a JEOL 2010 equipped with a high-brightness $LaB_6$ filament. Upon addition of base to the cationic solution of $Zn^{2+}$ and $Co^{2+}$, the color changed from the characteristic pink of octahedral $Co^{2+}$ to characteristic blue of tetrahedral $Co^{2+}$, shown in Fig. 1a. TEM images of the isolated nanocrystals show pseudo-spherical Wurtzite ZnO particles with diameters ranging from 3 to 6 nm depending on conditions (e.g., Fig. 1b). Electronic absorption spectra collected during base addition are presented in Fig. 1c. The bottom bold trace (red) shows the absorption spectrum prior to base addition, which exhibits only a weak absorption band centered at 19000 $cm^{-1}$. With base addition, intense new features appear at ~28000 and 17000 $cm^{-1}$ that are attributable to band gap and dopant-localized excitations of the product, $Co^{2+}$:ZnO. A broad absorption feature at ca. 18000 $cm^{-1}$ was also observed that is attributable to a tetrahedral $Co^{2+}$ intermediate. The isosbestic points in the ligand field region of Fig. 1c indicate that this intermediate is the direct precursor to the product, nanocrystalline $Co^{2+}$:ZnO. A more detailed discussion of these spectra is presented in reference 13.

Reaction yields were also examined as a function of initial $Co^{2+}$ concentrations in the ionic starting solutions. In these experiments, a sub-stoichiometric quantity of base (0.66 eq.) was added to the cationic solution. The system was allowed to reach pseudo-equilibrium conditions, after which the absorption spectrum was measured. This experiment was repeated at various doping levels, keeping all other experimental conditions constant. The results of these experiments are summarized in Fig. 2, which plots relative ZnO band gap absorption intensities vs. initial dopant concentrations. A very strong dependence of the ZnO yield on dopant concentration is evident, such that little or no ZnO was formed under these conditions at dopant concentrations above ca. 10%.

## Analysis and Discussion

For a fixed concentration of $Zn^{2+}$ ions, the intensity of the band gap absorbance under pseudo-equilibrium conditions provides a measure of the reaction yield for the synthesis of ZnO described by Equation 1.

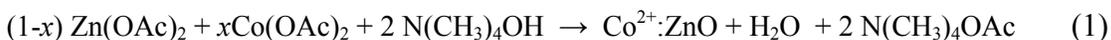

$$(1\text{-}x)\ \text{Zn(OAc)}_2 + x\text{Co(OAc)}_2 + 2\ \text{N(CH}_3)_4\text{OH} \ \rightarrow \ \text{Co}^{2+}\text{:ZnO} + \text{H}_2\text{O} + 2\ \text{N(CH}_3)_4\text{OAc} \qquad (1)$$

Figure 3 plots the intensity of the ZnO band gap absorbance from Fig. 1c as a function of added base. An induction volume of ca. 0.36 - 0.40 equivalents (eq.) of base is required before any ZnO nucleation is observed. At ca. 0.40 eq. of added base, a sharp jump in ZnO concentration is observed, indicating critical supersaturation of precursors has been exceeded. Subsequent base addition increases the ZnO concentration stoichiometrically, as seen from the linear dependence of ZnO absorbance on added OH⁻. Extrapolation of these data to zero intensity very nearly intercepts the origin, indicating that the vast majority of base is consumed to produce ZnO. These observations are generally consistent with the so-called LaMer model for crystallization,[19] in which precursor concentrations may be increased until a critical supersaturation concentration ($C_{ss}$) is reached, at which point crystallite nucleation occurs and the solution precursor concentration is depleted by diffusion-limited growth. This process is illustrated schematically in Fig. 4. Figure 4 describes an experiment in which $C_{ss}$ is reached only once, leading to a single nucleation event, but it is readily modified to accommodate serial nucleation events as in our base titration experiments.

The fate of cobalt throughout the synthesis is deduced from the $Co^{2+}$ ligand-field transitions observed at energies between 24 000 and 12 500 cm⁻¹ (Fig. 1c). Three distinct $Co^{2+}$ species are observed:[13] (1) the octahedral precursor (Figs. 1c, red), (2) a tetrahedral surface-bound $Co^{2+}$ intermediate whose intensities were deconvoluted by Single Value Decomposition and are shown in Fig. 3 (green), and (3) substitutional $Co^{2+}$ in Wurtzite ZnO (Fig. 1c, 3, blue). Little or no change in the $Co^{2+}$ absorption spectrum occurred until after ZnO nucleation was first observed (~0.36 equiv. of base). The titration data show formation of the tetrahedral surface-bound and substitutional $Co^{2+}$ species upon further base addition. The isosbestic points in Fig. 1c reveal that the surface-bound $Co^{2+}$ intermediate is the direct precursor to substitutional $Co^{2+}$ in ZnO, and hence that surface $Co^{2+}$ is internalized into the ZnO nanocrystal lattices during growth. Extrapolation of the substitutional $Co^{2+}$ absorption intensities to zero yields an x-axis intercept that coincides precisely with the point at which ZnO is first detected. In other words, $Co^{2+}$ ions are excluded from the initial nucleation event and are only incorporated into the ZnO lattice during growth, as illustrated schematically in Fig. 3(inset). $Co^{2+}$ incorporation during ZnO growth appears to be isotropic.[12]

The immediate precursors to ZnO nucleation are generally believed to be in the class of basic zinc acetate (BZA) clusters, which form with addition of base to solutions of $Zn(OAc)_2$. The smallest BZA cluster, $[(OAc)_6Zn_4O]$, has been isolated and studied crystallographically.[20] This cluster contains a recognizable subunit of the ZnO lattice, namely the $\mu_4$-oxo bridge of four $Zn^{2+}$ ions in a tetrahedral arrangement. In addition to the tetramer, a decamer $[(OAc)_{12}(Zn_{10}O_4)]$ has been detected by mass spectrometry in both ethanol[21] and DMSO[22] solutions of $Zn(OAc)_2$ after base addition. In the synthesis described here, the induction volume of base is thus likely consumed to form mixed-metal BZA clusters that serve as the immediate molecular precursors to $Co^{2+}$:ZnO nanocrystals. This assertion is supported by mass spectrometry data showing facile formation of $Co^{2+}/Zn^{2+}$ mixed-metal BZA clusters such as $Zn_3CoO(OAc)_9^+$ and $Zn_2Co_2O_2(OAc)_3^+$ in the gas phase from 1:1 mixtures of $Co(OAc)_2$ and $Zn(OAc)_2$.[23] From Fig. 1c, there is only a minor change in the $Co^{2+}$ ligand-field absorption band prior to ZnO nucleation, indicating that $Co^{2+}$ incorporated into the BZA clusters remains octahedrally coordinated.

Insight into the exclusion of $Co^{2+}$ from the initial ZnO nucleation event is gained by consideration of this phase transition in the context of the classical nucleation model. In this model, the driving force for the phase transition from solvated BZA molecular precursors to crystalline lattice is the substantial lattice energy of ZnO. This energy, parameterized by $\Delta F_v$, is the difference in the free energies of solvated and crystalline forms of the material. Its contribution to the total free energy of crystallization, $\Delta G$, increases with increasing crystal volume, $4/3\pi r^3$. For the precipitation of large crystals from solution, $\Delta F_v$ dominates the overall change in free energy. The initial product of homogeneous nucleation must be a small crystallite, however, which may subsequently grow into a large crystal. Small crystallites are destabilized by their surface free energies, $\gamma$, associated with their large surface areas and contributing to $\Delta G$ in proportion to the crystallite surface area, $4\pi r^2$. The overall free energy change for crystallization thus represents the competition between these two terms as described by Equation 2.[24]

$$\Delta G(r) = 4/3\,\pi r^3 \Delta F_v + 4\pi r^2 \gamma \tag{2}$$

Plotting Equation 2 as a function of crystallite radius, $r$, yields the reaction coordinate diagram shown in Fig. 5a. The maximum of this reaction coordinate diagram defines a critical radius, $r = r^*$, below which nuclei will redissolve ($dr/dt < 0$) and above which nuclei will survive to grow into observable crystals ($dr/dt > 0$). The activation barrier for this process is given by Equation 3, and defines the critical radius for nucleation, i.e. the smallest nucleus that can survive to develop into an observable particle.

$$\Delta G^* = \frac{16\pi\gamma^3}{3\Delta F_v^2} \tag{3}$$

Under diffusion limited growth conditions, the growth kinetics are described by Equations 4 and 5,[25]

$$\frac{dr}{dt} = \frac{K_D}{r}\left(\frac{1}{r^*} - \frac{1}{r}\right) \tag{4}$$

$$K_D = \frac{2\gamma D V_M^2 C_\infty}{RT} \tag{5}$$

where $V_M$ is the molar volume, D is the diffusion coefficient of the precursor, $C_\infty$ is the solubility of the solid with infinite dimension, R is the gas constant, and $T$ is the temperature. Equation 4 is plotted as a function of crystallite radius in Fig. 5b, and the volume of the growing crystallite is included in Fig. 5c for comparison.

Despite the very similar tetrahedral ionic radii of $Co^{2+}$ (~0.72 Å) and $Zn^{2+}$ (~0.74 Å),[26] substitution of $Co^{2+}$ to the ZnO lattice introduces strain that manifests itself as a dependence of the lattice parameters on dopant concentration (Vegard's law).[27] The dopant-induced strain in $Co^{2+}$:ZnO can be quantified by determination of the lattice constants $a_{Zn_{1-x}Co_xO}$ and $c_{Zn_{1-x}Co_xO}$ as a function of $x$. Although the thermodynamically favored lattice structure of CoO is the NaCl structure, Wurtzite CoO has been prepared[28] and provides a high-quality measure of $a$ and $c$ for $x = 1$. The data for $x = 0$ (Wurtzite ZnO)[29] and $x = 1$ (Wurtzite CoO)[28] are plotted in Fig. 6b and c, respectively, and Vegard's law indicates that the lattice parameter shifts for intermediate values of $x$ are expected to fall on this line. The shift in the *c* axis parameter (0.87% shift/mol fraction $Co^{2+}$) is indeed very close to the value of 0.7% shift/mol fraction of $Co^{2+}$ found experimentally for thin films of $Zn_{1-x}Co_xO$ ($0 < x < 0.20$)[30] prepared by pulsed laser deposition.[31]

In the terms of the classical nucleation model, dopant-induced lattice strain reduces the thermodynamic driving force for crystallization by diminishing $\Delta F_v$, making the doped crystallites slightly more soluble than their undoped counterparts. The decrease in $\Delta F_v$ in turn increases the activation barrier to crystallization, $\Delta G^*$, as shown in Fig. 5a(dotted). Furthermore, the plot in Fig. 5a also suggests that a decrease in $\Delta F_v$ necessitates a greater critical radius for doped particles (i.e., $r^*_{Co^{2+}:ZnO} > r^*_{ZnO}$), possibly placing greater demands on the nuclearity of the reaction. It is therefore more difficult to nucleate $Co^{2+}$:ZnO than it is to nucleate pure ZnO under identical conditions (*vide infra*). Importantly, although many BZA clusters contain dopants, at most typical doping concentrations there are always many BZA clusters that contain no dopants, and the system as a whole follows the least-energetic trajectory by nucleating pure ZnO. Dopants are then incorporated during the subsequent nanocrystal growth, where the driving force is large.

One of the manifestations of ZnO critical nucleus destabilization by $Co^{2+}$ is the inhibition of nucleation. Despite the very similar $Co^{2+}$ and $Zn^{2+}$ ionic radii, and the small ligand-field stabilization energy for octahedral $Co^{2+}$, the introduction of $Co^{2+}$ in the reaction mixture of Equation 1 has a large impact on the yield of the reaction. Reaction yields measured as a function of the initial dopant concentration (Fig. 2) reveal an exponential decrease in ZnO yield with increasing dopant concentrations, despite the fact that all concentrations studied are well below the solid solubility limit for $Co^{2+}$ in ZnO under these conditions. The disproportionately large inhibition of ZnO nucleation by $Co^{2+}$ is attributed to the inability of mixed-metal BZA clusters to nucleate the Wurtzite lattice structure because of their greater activation barrier (Fig. 5a). In this way, a single $Co^{2+}$ ion may render multiple $Zn^{2+}$ ions inert toward nucleation. From Fig. 2, addition of 2% $Co^{2+}$ prevents an average of 35% of the $Zn^{2+}$ ions from participating in nucleation, yielding an inhibition ratio of 18 $Zn^{2+}/Co^{2+}$. This ratio provides a crude but direct estimate of the size of the critical nucleus for ZnO in this reaction, i.e. only about 18 cations. This would correspond, for example, to the coalescence of only 2 decameric BZAs to form the critical nucleus of a ZnO nanocrystal.

The analysis of the critical nucleus size can be framed more quantitatively by considering the statistical distribution of dopants within BZA clusters of different sizes and at different doping levels. The statistics of BZA cluster doping are described by the binomial distribution in Equation 6, where $N$ is the total number of cation sites per cluster available for substitution and $n$ is the number of dopants per BZA cluster.

$$P(n|N) = \frac{(N)!}{n!(N-n)!}(xN)^n \big(1-(xN)\big)^{N-n} \tag{6}$$

For low doping concentrations, Equation 6 simplifies to the Poisson distribution function (Equation 7).

$$P(n) = \frac{(xN)^n e^{-xN}}{n!} \tag{7}$$

Figure 7a plots the probability distributions for doped BZA clusters of various nuclearities, calculated for $x = 0.02$ assuming a statistical distribution of dopants in the BZA clusters. As seen in Fig. 7a, the probability of having no dopants in a BZA cluster decreases with increasing cluster size. At a cluster size of 22 $Zn^{2+}$ cations, the calculated probability of having no dopants within the cluster is 65%, matching the experimental inhibition level for $x = 0.02$ from Fig. 2. Fitting the entire data set of Fig. 2 using Equation 7 yields an average cluster nuclearity of 25 ± 4 $Zn^{2+}$ ions (best fit = 24.8). The inhibition concentration dependence predicted from this cluster nuclearity is superimposed on the data in Fig. 2. For comparison, the calculated inhibition data for cluster sizes of 10 and 40 $Zn^{2+}$ ions are also included. Figure 7b shows the statistical distribution of dopants/cluster calculated for a 25 $Zn^{2+}$ BZA cluster at various dopant levels using Equation 7. The calculated probabilities closely reproduce the experimental nucleation inhibition data over the entire range of $Co^{2+}$ concentrations examined. The excellent agreement between the data and calculation strongly supports the interpretation that the inhibition arises from the incorporation of $Co^{2+}$ into BZA clusters, which increases their stability against nucleation of Wurtzite ZnO as described by Fig. 5a. These calculations also point to a critical nucleus of 25 ± 4 $Zn^{2+}$ ions for the nucleation of ZnO in DMSO.[22,32,33]

Using this estimate of the critical nucleus size and available calorimetric data, it is possible to estimate the increase in activation barrier that must be overcome for direct nucleation of doped ZnO from mixed-metal BZAs. The energy associated with dopant-induced strain in bulk $Co^{2+}$:ZnO can be assessed by determination of the enthalpy of formation, $\Delta H_f$, as a function of $Co^{2+}$ concentration, $x$. $\Delta H_f(x)$ values have been measured calorimetrically for $Zn_{1-x}Co_xO$ ($\Delta H_f(0.0337) = 23.6$ kcal/mol and $\Delta H_f(0.246) = 24.8$ kcal/mol) by dissolution in sulfuric acid (Equation 8),[34]

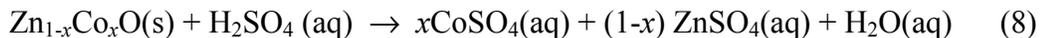

$$Zn_{1-x}Co_xO(s) + H_2SO_4\ (aq)\ \rightarrow\ xCoSO_4(aq) + (1-x)\ ZnSO_4(aq) + H_2O(aq) \qquad (8)$$

and are plotted in Fig. 6a relative to $\Delta H_f$ for pure ZnO (y-axis intercept). This plot yields the strain enthalpy per mol fraction of $Co^{2+}$ in ZnO and has a value of 5.74 kcal/mol/mol fraction $Co^{2+}$ in $Co^{2+}$:ZnO, indicating increasing lattice destabilization at higher $Co^{2+}$ concentrations. The relative stabilities deduced for $Co^{2+}$:ZnO dissolution in $H_2SO_4(aq)$ should hold for its synthesis in DMSO, with the equilibrium simply shifted to favor precipitation rather than dissolution. Approximating $\Delta S$ for Equation 8 to be independent of $x$ at these concentrations allows the association of $\Delta H_f$ with $\Delta F_v$ from Equation 2. From the above discussion, a BZA cluster containing 1 $Co^{2+}$ and 24 $Zn^{2+}$ ions will not nucleate, whereas an undoped cluster of 25 $Zn^{2+}$ ions will. By inspection of Fig. 6a, $\Delta H_f$ for this composition ($x = 0.04$) will be 0.23 kcal/mol ZnO less negative than for undoped ZnO, or 5.75 kcal/mol cluster. This difference is manifested in Equation 2 as an increase in the activation barrier, $\Delta G^*$, by 5.75 kcal/mol cluster. From Arrhenius expressions, this increase in $\Delta G^*$ leads to a rate constant for nucleation of a singly doped BZA cluster $1.5 \times 10^4$ times smaller than that of the undoped cluster.[35]

From the above analysis, we conclude that it is not possible to prepare $Co^{2+}$:ZnO nanocrystals with $Co^{2+}$ ions at their precise centers (i.e. within the critical nuclei) by homogeneous nucleation. This conclusion arises from very general physical considerations and is expected to also apply to other doped quantum dots (e.g. of CdS,

CdSe, etc.) as well. The generalization of this conclusion is supported by comparison of lattice strain data for $Co^{2+}$:ZnO with those of other lattices. Figures 6b and c show Vegard's law plots for solid solutions of $Cd_{1-x}Mn_xS$ and $Cd_{1-x}Mn_xSe$ obtained using experimental lattice parameters for the Wurtzite phases[29] of CdS/MnS and CdSe/MnSe. Solid solutions of these pairs are each known to obey Vegard's law even at dopant concentrations much larger than those typical of doped nanocrystals (e.g. up to $x \leq 0.5$).[3] The slopes for the *a* and *c* axes are 3.7 and 3.9 % shift/mol fraction Mn in $Cd_{1-x}Mn_xS$ and 4.2 and 4.3 % shift/mol fraction of Mn in $Cd_{1-x}Mn_xSe$, respectively. The lattice parameter shifts for $Cd_{1-x}Mn_xS$ and $Cd_{1-x}Mn_xSe$ are thus three to five times larger than the corresponding shifts of $Zn_{1-x}Co_xO$, from which we conclude that the strain induced by doping these lattices with 3d $TM^{2+}$ ions is even greater than that induced by doping ZnO with comparable ions. Incorporation of a $TM^{2+}$ impurity ion within the critical nucleus of a CdS or CdSe nanocrystal is therefore also less favorable than for ZnO. From this analysis we conclude that dopant exclusion from the critical nuclei of quantum dots prepared from solution is likely to be a general phenomenon.

In summary, we have analyzed the nucleation of $Co^{2+}$:ZnO nanocrystals from homogeneous solution in the context of a classical nucleation model. Addition of base to $Zn(OAc)_2$ solutions forms the well-known class of BZA clusters, which are the likely solution precursors in the nucleation of Wurtzite ZnO. Titration data revealed that $Co^{2+}$ impurities added to this reaction mixture are quantitatively excluded from the critical nuclei of ZnO but are readily incorporated into the nanocrystals during subsequent growth. The overall yield of ZnO nanocrystals was also reduced by introduction of $Co^{2+}$. These observations are explained by the conclusion that mixed-metal BZA clusters are unable to nucleate Wurtzite $Co_xZn_{1-x}O$. Instead, only nucleation involving BZA clusters containing no $Co^{2+}$ is observed. Statistical analysis shows that the occurrence of dopant-free BZA clusters diminishes exponentially with increasing dopant concentration, and the nucleation inhibition data are quantitatively reproduced for a critical cluster size of $25 \pm 4$ $Zn^{2+}$ ions. Using literature calorimetric data, the activation barrier for direct nucleation of $Co^{2+}$:ZnO containing only a single $Co^{2+}$ ion is 5.75 kcal/mol cluster greater than that for undoped ZnO, reducing the rate constant for nucleation of the former by a factor of over $10^4$. Such a strong effect is remarkable because of the very similar ionic radii and generally compatible chemistries of $Co^{2+}$ and $Zn^{2+}$, and this large effect emphasizes the extreme sensitivity of nucleation processes to minor perturbations. Consideration of data for other DMSs of which colloids have been investigated ($Mn^{2+}$:CdS and $Mn^{2+}$:CdSe) suggests that homogeneous nucleation of these DMSs will also likely suffer from the same basic physical restrictions. On the basis of these considerations, it appears unlikely that semiconductor quantum dots containing single dopants at their precise centers could be prepared by direct chemical routes, since critical nuclei that include dopants appear to be discriminated against kinetically.

**Acknowledgments.**

Financial support from the NSF (DMR-0239325) and ACS-PRF (37502-G) is gratefully acknowledged. D.R.G. is a Cottrell Scholar of the Research Corporation.

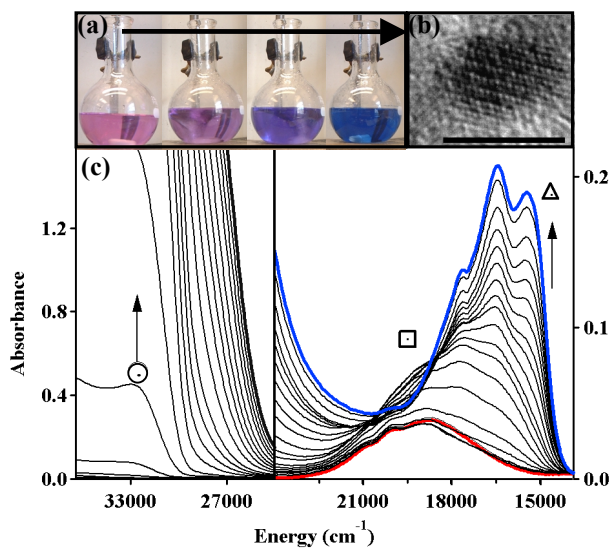

**Figure 1.** (color) (a) Photograph showing the change in color upon addition of OH⁻ to a solution of $Zn^{2+}$ and $Co^{2+}$. (b) Representative TEM image of a $Co^{2+}$:ZnO nanocrystal. Scale bar: 5 nm. (c) Electronic absorption spectra of the ZnO bandgap (left) and $Co^{2+}$ ligand field (right) energy regions collected as a function of added OH⁻, showing the band gap (○), intermediate $Co^{2+}$ (□), and substitutionally doped $Co^{2+}$:ZnO (Δ) spectroscopic features. Note the different scales. (Adapted from ref. 18)

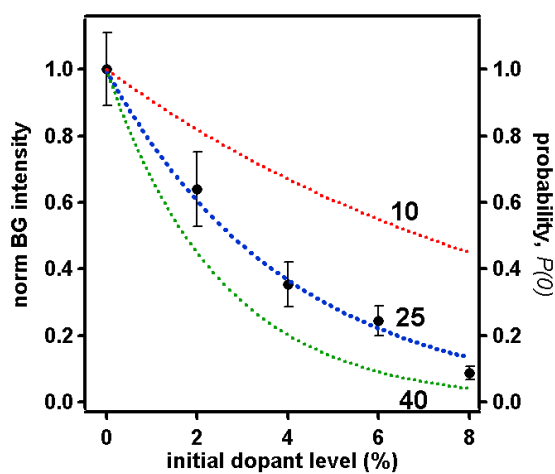

**Figure 2.** Relative band gap absorption intensity (●) plotted versus initial dopant level, measured after addition of 0.66 equivalents OH⁻ to a solution of $Zn(OAc)_2$ and $Co(OAc)_2$ at the indicated dopant level. The thick dotted line (blue) shows the statistical probability of a 25 $Zn^{2+}$ BZA cluster containing no $Co^{2+}$ ions (best fit=24.8), calculated from Equation 7 (see text). Calculated probabilities of 10 (red) and 40 (green) $Zn^{2+}$ BZA clusters containing no $Co^{2+}$ ions are included for comparison.

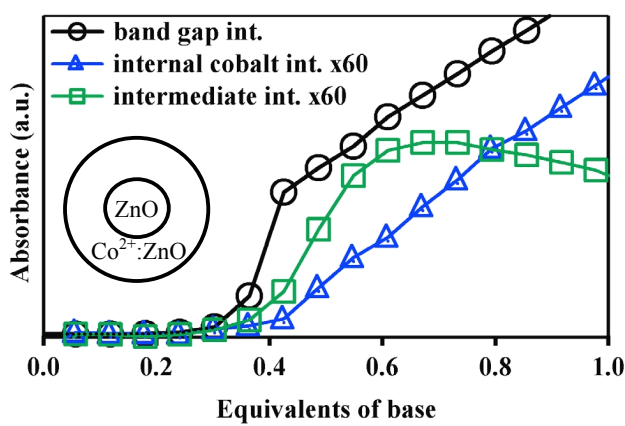

**Figure 3.** (color) ZnO bandgap (○), intermediate $Co^{2+}$ ligand-field (□), and $Co^{2+}$:ZnO ligand-field (△) absorption intensities deconvoluted from the data in Fig. 1 by Single Value Decomposition, plotted versus added base equivalents (Adapted from ref. 18). The inset depicts the existence of undoped ZnO cores in the nanocrystalline $Co^{2+}$:ZnO products.

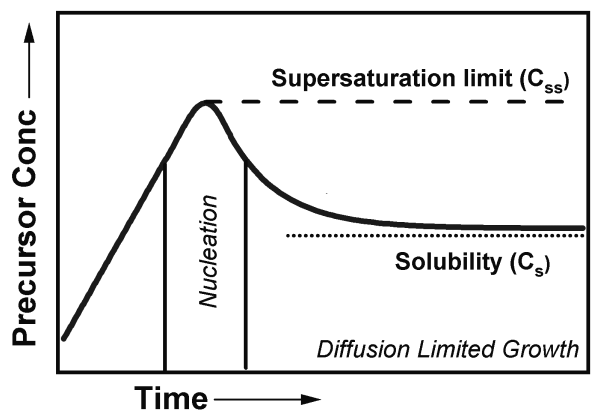

**Figure 4.** Reproduction of the generalized LaMer model plot describing nucleation and growth of crystallites versus time for continuous influx of precursor. Adapted from ref. 19.

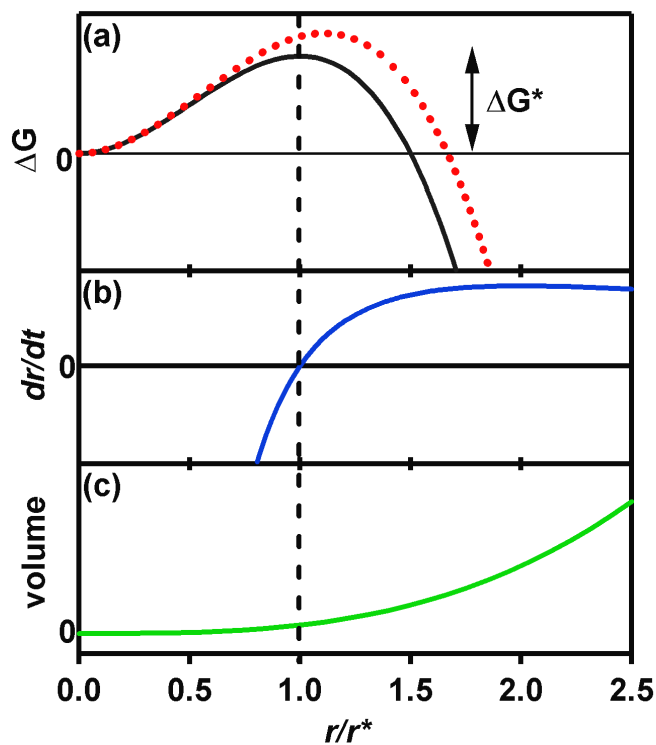

**Figure 5.** (color) (a) Gibbs free energy change (ΔG), (b) growth rate (*dr/dt*), and (c) crystallite volume plotted versus the crystallite radius, *r*, normalized by the critical radius, *r\**. The activation barrier to nucleation, $\Delta G^*$, determined from the classical nucleation model (Equation 2) is indicated in (a). The dotted line in (a) (red) shows the reaction coordinate diagram for nucleation including destabilizing impurity ions.

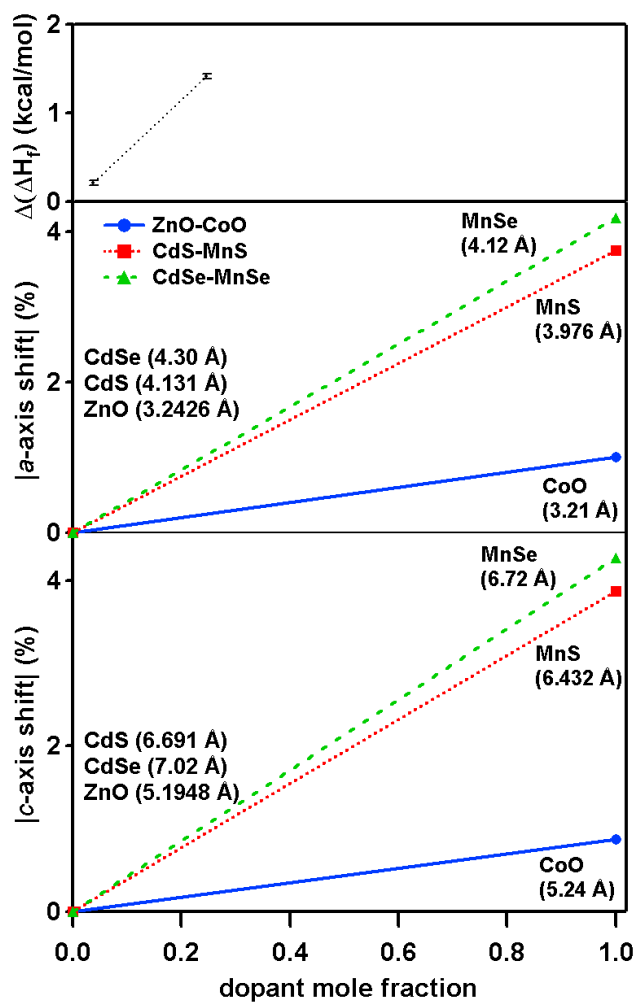

**Figure 6.** (color) (a) Enthalpy of formation of $Zn_{1-x}Co_xO$ measured by dissolution in $H_2SO_4$(aq) plotted versus *x*, absolute values of the relative (b) *a*-axis and (c) *c*-axis lattice parameter shifts for Wurzite phases of $Zn_{1-x}Co_xO$, $Cd_{1-x}Mn_xS$, and $Cd_{1-x}Mn_xSe$ DMSs. Shifts were calculated as percent of host lattice values. The lines represent solid solutions assuming adherence to Vegard's law. (Data taken from refs 28, 29, and 34)

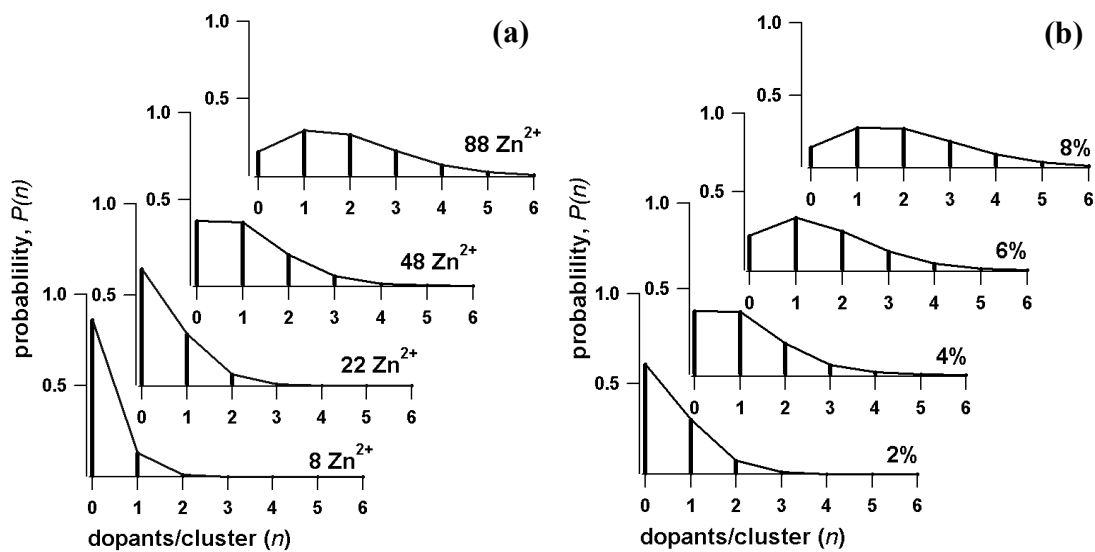

**Figure 7.** Poisson distributions showing the statistical doping probabilities at (a) various cluster sizes (indicated by the number of cations in the parent $Zn^{2+}$ cluster), calculated for 2% doping, and (b) various doping levels for BZA clusters containing 25 cations.

(1)     J. D. Bryan and D. R. Gamelin, Prog. Inorg. Chem. 54, 47, (2005); W. Chen, J. Z. Zhang and A. G. Joly, J. Nanosci. Nanotech., 4, 919 (2004).

(2)     S. A. Wolf, D. D. Awschalom, R. A. Buhrman, J. M. Daughton, S. von Molnár, M. L. Roukes, A. Y. Chthelkanova and D. M. Treger, Science, 294, 1488 (2001).

(3)     J. K. Furdyna, J. Appl. Phys., 64, R29 (1988).

(4)     J. K. Furdyna and J. Kossut In *Semiconductors and Semimetals*; R. K. Willardson and A. C. Beer, Eds.; Academic: N.Y., 1988; Vol. 25.

(5)     K. Ando In *Solid-State Sciences: Magneto-Optics*; Springer, (2000); Vol. 128, 211.

(6)     D. Ferrand, A. Wasiela, S. Tatarenko, J. Cibert, G. Richter, P. Grabs, G. Schmidt, L. W. Molenkamp and T. Dietl, Sol. State Comm., 119, 237 (2001).

(7)     Y. C. Tao, J. G. Hu and H. Liu, J. Appl. Phys., 96, 498 (2004); C. Gould, A. Slobodskyy, T. Slobodskyy, P. Grabs, C. R. Becker, G. Schmidt and L. W. Molenkamp, Phys. Stat. Sol. (b), 241, 700 (2004).

(8)     Y. Ohno, D. K. Young, B. Beschoten, F. Matsukura, H. Ohno and D. D. Awschalom, Nature, 402, 790 (1999); R. Fiederling, M. Keim, G. Reuscher, W. Ossau, G. Schmidt, A. Waag and L. W. Molenkamp, Nature, 402, 787 (1999); B. T. Jonker, Y. D. Park, B. R. Bennett, H. D. Cheong, G. Kioseoglou and A. Petrou, Phys. Rev. B, 62, 8180 (2000); V. N. Golovach and D. Loss, Semicond. Sci. Technol., 17, 355 (2002); H. Ohno, F. Matsukura and Y. Ohno, JSAP Int., 5, 4 (2002); C. Rüster, T. Borzenko, C. Gould, G. Schmidt, L. W. Molenkamp, X. Liu, T. H. Wojtowicz, J. K. Furdyna, Z. G. Yu and M. E. Flatté, Phys. Rev. Lett., 91, 216602 (2003); A. L. Efros, E. I. Rashba and M. Rosen, Phys. Rev. Lett., 87, 206601 (2001).

(9)     S. Chakrabarti, M. A. Holub, P. Bhattacharya, T. D. Mishima, M. B. Santos, M. B. Johnson and D. A. Blom, Nano Letters, ASAP (2005).

(10)    F. V. Mikulec, M. Kuno, M. Bennati, D. A. Hall, R. G. Griffin and M. G. Bawendi, J. Am. Chem. Soc., 122, 2532 (2000).

(11)    P. V. Radovanovic and D. R. Gamelin, J. Am. Chem. Soc., 123, 12207 (2001); N. S. Norberg, K. R. Kittilstved, J. E. Amonette, R. K. Kukkadapu, D. A. Schwartz and D. R. Gamelin, J. Am. Chem. Soc., 126, 9387 (2004).

(12)    P. V. Radovanovic, N. S. Norberg, K. E. McNally and D. R. Gamelin, J. Am. Chem. Soc., 124, 15192 (2002).

(13)    D. A. Schwartz, N. S. Norberg, Q. P. Nguyen, J. M. Parker and D. R. Gamelin, J. Am. Chem. Soc., 125, 13205 (2003).

(14)    J. Yin and Z. L. Wang, Handbook of Nanophase and Nanostructured Materials, 4, 174 (2003).

(15)    D. J. Norris, N. Yao, F. T. Charnock and T. A. Kennedy, Nano Lett., 1, 3 (2001).

(16)    A. P. Alivisatos, A. L. Harris, N. J. Levinos, M. L. Steigerwald and L. E. Brus, J. Chem. Phys., 89, 4001 (1988).

(17)    D. M. Hoffman, B. K. Meyer, A. I. Ekimov, I. A. Merkulov, A. L. Efros, M. Rosen, G. Counio, T. Gacoin and J.-P. Boilot, Sol. State Comm., 114, 547 (2000).

(18)    A. L. Efros and M. Rosen, Annu. Rev. Mater. Sci., 30, 475 (2000).

(19)    V. K. LaMer and R. H. Dinegar, J. Am. Chem. Soc., 72, 4847 (1950).

(20)    L. Hiltunen, M. Leskela, M. Makela and L. Niinisto, Acta. Chem. Scand. A, 41, 548 (1987).

(21)    M. Kohls, M. Bonanni, L. Spanhel, D. Su and M. Giersig, Appl. Phys. Lett., 81, 3858 (2002).

(22)    J. D. Bryan, K. Whitaker and D. R. Gamelin, to be published,  (2005).

(23)    G. C. DiDonato and K. L. Busch, Inorg. Chem., 25, 1551 (1986).

(24)    D. W. Oxtoby, Acc. Chem. Res., 31, 91 (1998).

(25)    T. Sugimoto, Adv. Colloid Interfac. Sci., 28, 65 (1987).

(26)    J. E. Huheey, E. A. Keiter and R. L. Keiter, *Inorganic Chemistry: Principles of Structure and Reactivity*, Harper Collins College Publishers, (1993).

(27)    L. Vegard and H. Schjelderup, Physik. Z., 18, 93 (1917).

(28)    M. J. Redman and E. G. Steward, Nature, 193, 867 (1962).

(29)    R. W. G. Wyckoff, *Crystal Structures*, Interscience Publishers, Inc., (1961).

(30)    H.-J. Lee, S.-Y. Jeong, C. R. Cho and C. H. Park, Appl. Phys. Lett., 81, 4020 (2002).

(31)    It is noted that a smaller shift is calculated (0.16 % shift/mol fraction $Co^{2+}$) using a very recently reported literature value for Wurtzite CoO, (A. S. Risbud, L. P. Snedeker, M. M. Elcombe, A. K. Cheetham and R. Seshadri, Chem. Mater., ASAP, (2005)).  Both of these calculations show the presence of a discernable lattice constant shift with increasing dopant concentration.

(32)    The same analysis of nucleation inhibition data obtained under identical conditions using $Ni^{2+}$ instead of $Co^{2+}$ (ref. 13) leads to an estimated critical cluster size of $14 \pm 2$ $Zn^{2+}$ ions. The difference between the critical cluster sizes determined from these two data sets may be due to non-statistical incorporation of $Ni^{2+}$ ions into BZA clusters because of the considerably smaller ligand substitution rates of octahedral $Ni^{2+}$. Further experiments are underway to understand this difference (ref. 22).

(33)    A cluster containing $25 \pm 4$ $Zn^{2+}$ cations would correspond to a pseudo-spherical Wurtzite ZnO nanocrystal of ca. $1.5 \pm 0.2$ nm diameter.  This critical radius size is similar to the radii of that of the smallest reported Wurtzite ZnO nanocrystals detected by X-ray diffraction, whose diameters were estimated using ultracentrifugation methods to be ~1.0 $\pm$ 0.1 nm (A. Wood, M. Giersig, M. Hilgendorff, A. Vilas-Campos, L. M. Liz-Marzan and P. Mulvaney, Aust. J. Chem., 56, 1051 (2003)). The difference between these two estimates is only ca. 5 $Zn^{2+}$ ions.

(34)    L. S. Kerova, M. P. Morozova and T. B. Shkurko, Russ. J. Phys. Chem., 47, 1653 (1973).

(35)    It is also possible that the doped crystallite would possess a greater surface free energy due to the same strain, causing further destabilization.  Indeed, ZnO nanocrystal growth is inhibited by surface-bound $Co^{2+}$ ions (ref. 13).